\documentstyle[bezier,12pt]{article}
\begin{document}
\begin{titlepage}
\hfill   SLAC-PUB-6496

\hfill   May 1994

\hfill   T/E
\vfill
\begin{center}
\bf\large
The Parton Distribution Functions in the Limit $x_{Bj}\to 1$
\end{center}
\vspace{0.25in}
\begin{center}
D. M\"uller\footnote{Supported in part by
                     Deutschen Akademischen Austauschdienst
              and by Department of Energy contract DE-AC03-76SF00515.}\\
\vspace{0.25in}
Stanford Linear Accelerator Center, \\
        P.O. Box 4349 Stanford, California 94309
\end{center}
\vspace{0.25in}
\begin{center}
{\it Presented at the Leipzig Workshop on }\\
{\it Quantum Field Theoretical Aspects of High Energy Physics,}\\
{\it Bad Frankenhausen, Germany,September 20-24, 1993.}
\end{center}
\vfill
\begin{abstract}
We review both the counting rule and the influence of the
evolution in $Q^2$ for the large $x_{Bj}$ behaviour of the valance quark
distribution functions.
Based on a factorization procedure we present a more general perturbative
treatment to compute this behaviour.
A complete analysis is performed in the scalar  $\phi^3_{[6]}$-theory for the
parton distribution function of the ``meson'',
which shows that logarithmical corrections arise from the distribution
amplitude and that the reference momentum square $Q_0^2$ is fixed by
$x_{Bj}$.
\end{abstract}
\vspace{0.25in}
\begin{center}
\mbox{\ }
\end{center}
\vfill
\end{titlepage}
\newpage

\section{Introduction}

Deep inelastic lepton-hadron scattering (DIS) experiments have shown that
hadrons consist of point-like particles called partons.
These partons were identified as quarks and gluons, which are the fundamental
degrees of freedom  of a relativistic $\mbox{SU}_c(3)$ gauge field theory,
quantum chromodynamics (QCD).
For instance, corresponding to this quark-parton model a nucleon is formed by
a colour singlet state that consists of three valence quarks as well as
sea quarks and gluons.

Since the bound state problem remains unsolved at present it is not possible
to compute the DIS cross section completely from the first principle of QCD.
However, with the help of the operator product expansion \cite{OPE}
it could be shown that
the hadronic part factorises as a convolution of perturbative
computable Wilson coefficients and nonperturbative parton
distribution functions \cite{OPE-DIS}.\footnote{At first this factorization was
performed in terms of the moments  (expectation values of local operators) from
which the mentioned convolution can be obtained with the help of the Mellin
transformation.}
These distribution functions are defined as forward matrix elements of string
light-cone (LC) operators \cite{BroLep80,ColQiu89}.
Moreover, from the renormalization group  (RG) equation of these operators one
obtains the Gribov-Lipatov-Altarelli-Parisi (GLAP) equation
\cite{AltPar77Lip74},
which controls the evolution of the distribution functions in $Q$.
However, it remains a theoretical non-predictable input, the distribution
functions at a reference momentum $Q_0 \sim 1$ GeV.

At large $x_{Bj}$ the theoretical situation is much better.
Therefore, it is possible,  with the help of the naive quark-parton model,
to derive a counting rule \cite{allg.CR}
that predicts the power behaviour of the distribution functions in
$(1-x_{Bj})$ for both mesons and baryons \cite{x-1CR}.
These predictions will be reviewed in Section 2.

In the Section 3 we define the distribution functions as expectation
values of LC operators and discuss, with the help of their RG equation (or
GLAP equation), the change of the large $x_{Bj}$ behaviour by evolution in
$Q$.

In Section 4 the distribution functions will be factorized as a convolution of
distribution amplitudes and a perturbative computable hard amplitude,
which is computed for the ``meson'' distribution function of the scalar
$\phi^3_{[6]}$-theory in six dimensions.
Apart from the leading power behaviour the convolution also provides
logarithmical correction in $(1-x_{Bj})$ that arises from the evolution
of the distribution amplitude, and also the value of the
reference momentum square $Q_0$ as a function of $x_{Bj}$.

\section{Counting rule}

It is well known that the hadronic part of the unpolarized charged lepton DIS
is described by the hadronic tensor
\begin{eqnarray}
W_{\mu\nu}(x_{Bj},Q^2) &=& \left(-g_{\mu\nu} + {q_\mu q_\nu \over q^2}\right)
F_1(x_{Bj},Q^2)
\nonumber\\
                        & &+\left(P_\mu-{q_\mu Pq \over q^2}\right)
\left(P_\nu-{q_\nu Pq \over q^2}\right) F_2(x_{Bj},Q^2)/Pq,
\end{eqnarray}
where $P$ is the four-momentum of the hadron, and $q$ denotes the four-momentum
transfer.
Here $F_1(x_{Bj},Q^2)$ and $F_2(x_{Bj},Q^2)$ are the structure functions
which depend on the Bjorken variable $0\le x_{Bj}  = -q^2/2qP\le 1$ and
the momentum transfer $Q^2= -q^2$.

In the Bjorken region, i.e., large momentum transfer $Q$ and fixed $ x_{Bj}$,
the two structure functions can be expressed according to the quark-parton
model by the non-perturbative quark distribution functions
$q_i(x_{Bj},Q^2)$ and anti-quark distribution functions:
$\bar{q}_i(x_{Bj},Q^2)$,
\begin{eqnarray}
\label{GS-relation}
F_1(x_{Bj},Q^2) = F_2(x_{Bj},Q^2)/(2x_{Bj})
                = {1\over 2}\sum_{i}^{} e_i^2
            \left(q_i(x_{Bj},Q^2)+ \bar{q}_i(x_{Bj},Q^2)\right),
\end{eqnarray}
where $i=u,d,...$ is the flavor index, and $e_i$ is the corresponding
electrical charge.

As mentioned, at large $x_{Bj}$ it is possible to compute the parton
distribution functions.
But one deals with two large scales, $Q\to\infty$ and $1/(1-x_{Bj})\to\infty$,
so one has to clarify in which way the two limits will be performed.
With the help of the hadronic final state mass
$M_x^2 = (P+q)^2 = M_H^2 + (1-x_{Bj})Q^2/ x_{Bj}  \ge M_H^2$, where  $M_H$ is
the mass of the hadron, one finds the following classification:
\begin{enumerate}
\item
$(1-x_{Bj})Q^2 \to 0$, \ \ \ \ \ \ \ \ \ $M_x^2=M_H^2$,
\item
$(1-x_{Bj})Q^2 \to \mbox{finite}$, \ \ \ \ $M_x^2=M_H^2+\mbox{finite}$,
\item
$(1-x_{Bj})Q^2 \to \infty $, \ \ \ \ \ \ \ $M_x^2 \to \infty$.
\end{enumerate}
In this paper we consider only the third case, for which the factorization in
$Q$ via the operator product expansion remains true so that, as usual, higher
order twist terms can be neglected.

Now let us review the counting rule for the large $x_{Bj}$ behaviour of the
structure function.
Corresponding to the naive quark-parton model we have the struck quark with
momentum $p$, and the spectator quarks carry the momentum $P-p$:
\begin{eqnarray}
\label{impulse}
p = \left(x_{Bj}P_+,{M^2+\vec{k}^2_\perp \over x_{Bj}P_+
},\vec{k}_\perp\right),\qquad
P-p = \left((1-x_{Bj})P_+,{{\cal M}^2+\vec{k}^2_\perp \over (1- x_{Bj})P_+}
               ,-\vec{k}_\perp \right).
\end{eqnarray}
Here we have introduced light-cone coordinates, i.e.,
$p = (p_+=p_0+p_3,p_-=p_0-p_3,\vec{p}_\perp)$,
so that $x_{Bj}= p_+ /P_+$.
The essential assumption for deriving the counting rule is the requirement that
the spectator quarks have a non-vanished finite mass $\cal{M}$.
Then from the kinematical identity for the hadron mass $M^2_H=((P-p) + p)^2$
it follows that for $x_{Bj}\to 1$ the mass of the struck quark
\begin{eqnarray}
\label{m-strqua}
M^2=p^2= {x_{Bj}(1-x_{Bj})M^2 - x_{Bj}
                  {\cal M}^2 -\vec{k}_\perp^2 \over 1-x_{Bj}}
     \sim -{{\cal M}^2 +\vec{k}_\perp^2 \over 1-x_{Bj}}
\end{eqnarray}
is far off shell.
Note that this behaviour also remains in the infinite momentum frame.

The power behaviour of the structure functions at $x_{Bj}\to 1$ follows
directly from the far off shell behaviour of the wave function.
The iteration of the Bethe-Salpeter equation provides that this behaviour
is determined by the minimal number of gluon exchanges required
to stop the spectator  (see Fig.\ 1 for a nucleon).
Thus, without taking into account the spin of the partons one gets the
following counting rule \cite{x-1CR}:
\begin{eqnarray}
\label{countrule}
F_2(x_{Bj}) \sim q(x_{Bj})
            \sim  (1-x_{Bj})^{2n_s-1}\quad\mbox{for\ }x_{Bj}\to1,
\end{eqnarray}
where $n_s$ is the number of spectators.

Corresponding to the counting rule (\ref{countrule}) a nucleon has a power
behaviour of 3, which is consistent with the experimental results.
A meson should have a power behaviour of 1.
But it was shown in Ref.\ \cite{Eza74} that spin effects provide an additional
suppression so that the power is 2.
It is interesting to note that experimentally one finds for the pion the
power behaviour is closer to 1 than to 2 \cite{expVerh};
so that the situation about this prediction is not quite clear
(see also Ref.\ \cite{x-1CR} ).

\section{Improving the counting rule result by evolution}

The predicted behaviour of the structure function at large $x_{Bj}$ will be
modified by the evolution of the distribution functions and by perturbative
corrections of the Wilson coefficients.
Although at  $x_{Bj} \to 1$ higher order corrections are important
\cite{NeeZij92}, in this paper we discuss only the modification by the
evolution at leading order.
Interesting results from perturbative all-order investigations can be found in
the literature for both the evolution of the distribution functions
\cite{KorMar93} and corrections of the Wilson coefficients \cite{Ste87}.

First let us sketch why the distribution functions are defined as
expectation values of LC operators and why the GLAP equation arises from the
RG equation of these operators.
The dispersion relation
\begin{eqnarray}
\label{disperrel}
W_{\mu\nu}(x_{Bj},Q^2)&=&{1 \over 4\pi} \mbox{Im} T_{\mu\nu}(x_{Bj},Q^2),
\nonumber\\
T_{\mu\nu}(x_{Bj},Q^2)&=&i\int d^4x <P|\mbox{T}\{j_\mu(x)j_\nu(0)\}|P>e^{iqx},
\end{eqnarray}
provides that the hadronic tensor can be expressed by the imaginary part of the
forward compton scattering amplitude $T_{\mu\nu}(x_{Bj},Q^2)$.
Furthermore, the leading $Q^2$ behaviour of $W_{\mu\nu}(x_{Bj},Q^2)$ is
determined by the light-cone singularities of the time ordered product
$\mbox{T}\{j_\mu(x) j_\nu(0)\}$.
Thus, the non-local light-cone expansion of this operator product
\cite{AniZav78}
provides a definition of the quark distribution functions in terms of gauge
invariant string operators that are a resummation of local leading twist
operators (see, e.g., \cite{BraGeyHorRob87}).
In the axial gauge $\tilde{x}A=0$ these operators are simplified to be
bi-local:
\begin{eqnarray}
\label{}
q_i(x,Q^2)&=& \int {d\kappa \over 2\pi} e^{2i\kappa \left(\tilde{x}P\right) x}
<P|\bar{\psi}_i(-\kappa \tilde{x}) \tilde{x}\gamma
                                    \psi_i(\kappa \tilde{x})|P>_{|\mu^2=Q^2},
\mbox{\ } 0\le x\le 1,\\
\bar{q}_i(x,Q^2)&=& -\int {d\kappa \over 2\pi}
                     e^{-2i\kappa\left(\tilde{x}P\right) x}
<P|\bar{\psi}_i(-\kappa \tilde{x}) \tilde{x}\gamma
                                    \psi_i(\kappa \tilde{x})|P>_{|\mu^2=Q^2},
\mbox{\ } 0\le x\le 1.
\end{eqnarray}
Here the more arbitrary light-like vector $\tilde{x}$ arises from the
projection of $x$ onto the light-cone and can be chosen as
$\tilde{x}=(\tilde{x}_+=0,\tilde{x}_-=2,\vec{0}_\perp)$
so that $\tilde{x}P= P_+$.

In the following we consider only the evolution in the flavor non-singlet
sector.
For technical simplification we define formally the flavor non-singlet
operators
\begin{eqnarray}
\label{defoper}
 O^a(x,\mu^2) = \int {d\kappa \over 2\pi} e^{2i\kappa \left(\tilde{x}P\right)x}
\bar{\psi}_i(-\kappa \tilde{x}) \tilde{x}\gamma \lambda_{ij}^a \psi_i(\kappa
\tilde{x}),
\end{eqnarray}
where $\lambda_{ij}^a$ is a generator of the flavor group.
If we keep in mind the forward case then the renormalization group equation
can be formally written as
\begin{eqnarray}
\label{RGE}
\mu {d^2 \over d\mu^2} O^a(x,\mu^2) = \int_x^1 {dy \over y}
P(x/y;\alpha_s(\mu^2)) O^a(y,\mu^2),
\end{eqnarray}
where the integral kernel is known as perturbative expansion in $\alpha_s$,
\begin{eqnarray}
\label{GLAP-kernel}
P(x;\alpha_s)= C_F {\alpha_s \over 2 \pi}\left({1+x^2 \over (1-x)_+}+{3\over 2}
\delta(1-x)\right)+ O(\alpha_s^2),\quad \mbox{\ where\ } C_F=4/3.
\end{eqnarray}

Forming forward matrix elements and setting the renormalization point $\mu = Q$
one gets from the RG equation (\ref{RGE}) the GLAP equation.
Take into account Eq.\ (\ref{GLAP-kernel}) one finds for $x\to 1$
\begin{eqnarray}
Q^2 {d \over dQ^2} q^{NS}(x,Q^2) &=& C_F {\alpha_s(Q^2) \over 2\pi}
\Bigg(\left(2\ln(1-x) -{3 \over 2} \right)q^{NS}(x,Q^2)
\nonumber\\
& & + 2\int_{x}^{1}dy {q^{NS}(y,Q^2)-q^{NS}(x,Q^2) \over y-x}\Bigg).
\end{eqnarray}
This equation is to be implemented by a initial condition that has,
corresponding to Eq.\ (\ref{countrule}), the form
$q^{NS}(x,Q_0^2) \propto (1-x)^\nu$, where the value of $\nu$ is approximately
determined by the counting rule result.
Using the ansatz
$q^{NS}(x,Q^2) = A(Q,Q_0) (1-x)^{\nu+\xi(Q,Q_0)}$, where $\xi(Q_0,Q_0)=0$,
it is straight forward to derive the solution \cite{Dok78}
\begin{eqnarray}
\label{beh.ofDA}
q^{NS}(x,Q^2) \propto {\Gamma(\nu+1) \over \Gamma(\nu+1+\xi(Q,Q_0))}
          e^{(3/4-\gamma_E)\xi(Q,Q_0)} (1-x)^{\nu+\xi(Q,Q_0)},
\end{eqnarray}
where
\begin{eqnarray}
\label{}
\xi(Q,Q_0) = 2C_F \int_{Q_0^2}^{Q^2}{dt \over t} {\alpha_s(t) \over 2\pi}
           = {4 C_F \over \beta_0} \ln\left({\ln Q^2/\Lambda^2 \over \ln
Q^2_0/\Lambda^2}\right),\quad \beta_0 = 11-(2/3)n_f,
\end{eqnarray}
is a function increasing with $Q$.
Thus, with rising $Q$ the $(1-x)$ power behaviour increases and the prefactor
decreases.

\section{Factorization of the distribution function}

In this section we present a more rigorous treatment that allows the
perturbative computation of the large $x_{Bj}$ behaviour of the parton
distribution functions and also provides a smooth inclusive exclusive
connection.
As we saw in Section 2, at $x_{Bj} \to 1$, the parton model suggests that the
momentum of the struck quark is far off shell.
Thus, using the light-cone quantization and the representation of the wave
function by a superposition of fock states, a factorization can be performed
based on the iteration of the equation of motion \cite{BroLep80},\footnote{A
similar treatment is used for the factorization of exclusive large momentum
transfer processes.}
\begin{eqnarray}
\label{fac-qdf}
q(x,Q^2) \sim \int_0^1 du\int_0^1 dv
            \phi^\ast(u,\lambda^2) T_H(x,u,v,\lambda^2,Q^2) \phi(v,\lambda^2),
\end{eqnarray}
where the factorization scale is given by $\lambda^2\sim m^2 / (1-x)$.
Here the hard amplitude  $T_H(x,u,v,\lambda^2,Q^2)$ includes the
LC operator $O^a(x,\mu=Q)$ and provides, in principal (see below), the power
behaviour in $(1-x)$.

For simplicity, in the following we restrict ourselves to the parton
distribution functions of the meson.
Since the large momentum arises from the kinematics of the partons inside the
meson, the leading order approximation of $T_H(x,u,v,\lambda^2,Q^2)$ is
determined by the two gluon exchanges (see Fig.\ 2).
\begin{figure}[h]
\unitlength1mm
\hfil
\begin{minipage}[b]{70mm}
\unitlength1mm
\begin{picture}(38.00,17.00)(-16,0)
\thicklines
\put(10.00,10.00){\circle{10.00}}
\put(10.00,10.00){\makebox(0,0)[cc]{$P$}}
\put(5.00,11.00){\line(-1,0){8.00}}
\put(5.00,9.00){\line(-1,0){8.00}}
\put(3.00,10.00){\line(-3,2){3.6}}
\put(3.00,10.00){\line(-3,-2){3.6}}

\put(13.00,6.00){\line(1,0){20}}
\put(13.00,14.00){\line(1,0){20.00}}
\put(15.00,10.00){\line(1,0){18.00}}
\bezier{6}(20.00,14.00)(20.00,12.00)(20.00,10.00)
\bezier{6}(25.00,10.00)(25.00,8.00)(25.00,6.00)
\put(20.00,14.00){\circle*{0.8}}
\put(20.00,10.00){\circle*{0.80}}
\put(25.00,10.00){\circle*{0.80}}
\put(25.00,6.00){\circle*{0.80}}
\end{picture}\par
\parbox[t]{70mm}{\footnotesize Fig.\ 1:  This graph yields to the leading
  $(1-x)^3$ power behaviour of the nucleon structure function for $x\to 1$.}
\end{minipage}
\hfil
\unitlength1mm
\begin{minipage}[b]{80mm}
\unitlength1mm
\begin{picture}(66.00,20.00)(-3,0)
\thicklines
\put(10.00,10.00){\circle{10.00}}
\put(60.00,10.00){\circle{10.00}}

\put(13.00,6.00){\vector(1,0){7.5}}
\put(20.50,6.00){\vector(1,0){15.50}}
\put(36.0,6.00){\vector(1,0){15.50}}
\put(51.50,6.00){\line(1,0){5.50}}

\put(13.00,14.00){\vector(1,0){7.50}}
\put(20.50,14.00){\vector(1,0){9.0}}
\put(29.50,14.00){\line(1,0){2.5}}

\put(38.00,14.00){\vector(1,0){4.50}}
\put(42.50,14.00){\vector(1,0){9.0}}
\put(51.50,14.00){\line(1,0){5.50}}

\put(19.50,16.00){\makebox(0,0)[cb]{$\scriptstyle p_1$}}
\put(50.50,16.00){\makebox(0,0)[cb]{$\scriptstyle p_2$}}
\put(28.50,16.00){\makebox(0,0)[cb]{$\scriptstyle k$}}
\put(41.50,16.00){\makebox(0,0)[cb]{$\scriptstyle k$}}
\put(35.0,4.00){\makebox(0,0)[ct]{$\scriptstyle P-k$}}
\put(19.50,4.00){\makebox(0,0)[ct]{$\scriptstyle P-p_1$}}
\put(50.50,4.00){\makebox(0,0)[ct]{$\scriptstyle P-p_2$}}

\bezier{13}(25.00,14.00)(25.00,10.00)(25.00,6.00)
\bezier{13}(45.00,14.00)(45.00,10.00)(45.00,6.00)
\put(25.00,14.00){\circle*{0.80}}
\put(25.00,6.00){\circle*{0.80}}
\put(45.00,14.00){\circle*{0.80}}
\put(45.00,6.00){\circle*{0.80}}
\put(5.00,11.00){\line(-1,0){8.00}}
\put(5.00,9.00){\line(-1,0){8.00}}
\put(65.00,11.00){\line(1,0){8.00}}
\put(65.00,9.00){\line(1,0){8.00}}
\put(70.00,10.00){\line(-3,-2){3.6}}
\put(70.00,10.00){\line(-3,2){3.63}}
\put(3.00,10.00){\line(-3,2){3.6}}
\put(3.00,10.00){\line(-3,-2){3.6}}

\put(35.00,14.00){\circle{6.00}}
\put(33.00,12.00){\line(1,1){4}}
\put(33.00,16.00){\line(1,-1){4}}

\put(10.00,10.00){\makebox(0,0)[cc]{$P$}}
\put(60.00,10.00){\makebox(0,0)[cc]{$P$}}
\end{picture}\par
\parbox[t]{80mm}{\footnotesize Fig.\ 2: Parton distribution function of
        the meson at large $x$.
        The crossed circle symbolizes the LC operator insertion.}
\end{minipage}
\hfil
\end{figure}
Note that $T_H$ is not dependent on the reference momentum $Q_0$ and that
the $Q$-dependence will be induced by the renormalization of the LC operator.
Furthermore, the meson distribution amplitude $\phi(u,\lambda^2)$ satisfies an
evolution equation which controls the logarithmical corrections in
$\lambda^2$ \cite{BroLep80}:
\begin{eqnarray}
\phi(u,\lambda^2) = \sum_{k=0}^{\infty} a_k (1-u)u C_k^{3/2}(2u-1)
 \left({\ln(\lambda^2/\Lambda^2) \over
\ln(\lambda_0^2/\Lambda^2)}\right)^{\gamma_k^0/\beta_0},
\end{eqnarray}
where
$\gamma_k^{0} = C_F\left(3+{2\over (k+1)(k+2)} - 4\sum^{k+1}_{i=1} {1\over
i}\right)$,
$C_k^{3/2}$ are Gegenbauer polynomials, $a_k$ are nonperturbative
coefficients,
and $\lambda_0^2= m^2/(1-x_0)$ is an appropriate reference momentum square.

Now we study the large $x$ behaviour of the parton distribution functions for
the scalar $\phi^3_{[6]}$-theory.
First we compute the hard amplitude  $T_H$ by including the tree
approximation of the LC operator vertex, which is given by $\delta(x-k_+/P_+)$.
Using the far off shell behaviour of the struck partons
[see Eq.\ (\ref{m-strqua})]
one finds corresponding to Fig.\ 2 (substitute the gluon lines by scalar lines)
the following Feynman integral representation,
\begin{eqnarray}
\label{intrepr}
T_H(x,u,v)
 \sim\int_{0}^{1}dx_1 \int_{0}^{1-x_1}dx_2 \int_{0}^{1-x_1-x_2} dx_3
{1-x_1-x_2-x_3 \over (D-B^2)^2} \delta(x - x_1 u -x_2 v -x_3),
\nonumber\\
D-B^2\sim -(1-x_1-x_2-x_3)\left(x_1{m^2\over 1-u}+x_2{m^2\over 1-v}
                          -M^2_H\right)  - [1-(x_1+x_2)x_3] m^2,
\end{eqnarray}
where the mass of the spectator partons was set equal to the free particle
mass $m$.
The restriction of the integration region by the $\delta$-function comes from
the operator vertex and provides three contributions,
\begin{eqnarray}
\label{fac-th1}
T_H(x,u,v) &=& \theta\left(u,v<x\right) T_H^1 +
 \left\{\theta\left(u<x<v\right) T_H^{2}+\{u\leftrightarrow v\}\right\}
                +\theta\left(x<u,v\right) T_H^3,
\nonumber\\
T_H^1(x,u,v) &=& (1-x)\left({1-x \over 1-u}\right) \left({1-x \over 1-v}\right)
f_1(x,u,v),
\nonumber\\
T_H^{2}(x,u,v) &=& (1-x) \left(  {1-x \over 1-u} f_2(x,u,v) +
                               {v-x \over v-u} \tilde{f}_2(x,u,v)\right),
\nonumber\\
T_H^3(x,u,v) &=& (1-x) f_3(x,u,v),
\end{eqnarray}
where $f_i$,\ $i= 1,2,3,$ and $\tilde{f}_2$ are smooth functions,
which can be more or less replaced by constants.

The convolution (\ref{fac-qdf}) provides the $x\to 1$ behaviour; but
because of the complicated structure of $T_H(x,u,v)$ in Eq.\ (\ref{fac-th1})
the result depends on the end point behaviour of the
distribution function $\phi(u)$.
In the physical case, where the distribution amplitude $\phi(u)$ vanishes at
the end points, the leading term comes from the integration region
$u,v\le x \sim 1$, which provides\footnote{Because of the space-time dimension
six the obtained power behaviour is 3 and it agrees with the counting rule
result.}

\begin{eqnarray}
\label{res-ph}
q(x)=
\int_0^x du\int_0^x dv \phi^\ast(u,\lambda^2) T_H^1(x,u,v) \phi(v,\lambda^2)
          \sim (1-x)^3 \left|\int_{0}^{1}du {\phi(u)\over 1-u}\right|^2
\quad\mbox{for\ }x\to 1.
\end{eqnarray}

It is also interesting to investigate the unphysical case, in which the
distribution amplitude is constant or divergent at the end points,
i.e.,\ $\phi(u) \sim (1-u)^{-\epsilon}$ for $u\to 1$, $\epsilon \ge 0$.
Then one gets for $1/2 > \epsilon$
\footnote{This restriction is necessary to avoid divergencies caused by the
convolution.}
the behaviour
\begin{eqnarray}
\label{}
q(x)\sim (1-x)^3 |\phi(x)|^2 \sim (1-x)^{(3-2\epsilon)}.
\end{eqnarray}
Obviously, the obtained power behaviour is different from
that given in Eq.\ (\ref{res-ph}).

To take into account the evolution in $Q$ we
determine $O(x,\mu^2/k^2,\alpha_s(\mu))$ for $x\to 1$ from the RGE
and include it in the computation of $T_H$.
For $x\to 1$ the integral kernel  is completely determined by the anomalous
dimension $\gamma_\phi(\alpha_s)$ of the field $\phi$,
$P(x;\alpha_s) =
(\alpha_s/(2\pi)) \gamma_\phi^0 \delta(1-x) + O(\mbox{const}.,\alpha_s^2)$.
Thus, the RGE is simplified to
\begin{eqnarray}
\label{}
\mu^2 {d \over d\mu^2} O(x,\mu^2/k^2,\alpha_s(\mu))
\begin{array}{c}
\mbox{}\\{\displaystyle =} \\{\scriptstyle x\to 1}
\end{array}
 {\alpha_s(\mu^2)\over 2\pi} \gamma_\phi^0 O(x,\mu^2/k^2,\alpha_s(\mu)).
\end{eqnarray}
The solution of this first order (partial) differential equation can be found
in a text book:
\begin{eqnarray}
\label{RGESol}
O(x,\mu^2/k^2,\alpha_s(\mu))
&=&
                              O(x,1,\alpha_s(k^2))
                     \left({\ln(\mu^2/\Lambda^2)\over
                     \ln(k^2/\Lambda^2)}\right)^{2\gamma^0_\phi/\beta_0},
\nonumber \\
&=&
         \left(\delta\left(x-{k_+ \over P_+ }\right) + O(\alpha_s(k^2))\right)
         \left({\ln(\mu^2/\Lambda^2)\over
         \ln(k^2/\Lambda^2)}\right)^{2\gamma^0_\phi/\beta_0}.
\end{eqnarray}
Here $O(x,1,\alpha_s(k^2))$ represents the initial condition, which,
assuming $\alpha_s(k^2)$ is small enough, can be expanded perturbatively.

Including the solution (\ref{RGESol}) in the computation of  $T_H$ and setting
the renormalization point $\mu= Q$ provides for a physical distribution
amplitude:
\begin{eqnarray}
\label{}
q(x,Q^2)
\begin{array}{c}
\mbox{}\\ {\displaystyle \sim} \\{\scriptstyle x\to 1}
\end{array}
          \left({\ln(Q^2/\Lambda^2)\over
          \ln(m^2/((1-x)\Lambda^2)}\right)^{2\gamma^0_\phi/\beta_0}
       (1-x)^3 \left|\int_{0}^{1}du {\phi(u,m^2/(1-x))\over 1-u}\right|^2.
\end{eqnarray}
The comparison with the solution of the GLAP equation shows that the
reference momentum square $Q_0^2 = m^2/(1-x)$.

\section{Summary}

Using both the definition of the parton distribution functions as forward
matrix elements of light-cone operators and the parton model describing the
hadronic bound state we have factorized
(for $Q^2 > m^2/(1-x_{Bj})$ and large $m^2/(1-x_{Bj})$)
this distribution function as a convolution of a hard amplitude
and a distribution amplitude.
Here the factorization scale $\lambda\sim m^2/(1-x_{Bj})$ originates from
the kinematics of the parton inside the hadron as well as the assumption that
the spectator mass does not vanish.
Furthermore, we have computed the hard amplitude in the scalar
$\phi^3_{[6]}$ toy theory for the distribution function of the ``meson'' and
found that the resulting $(1-x_{Bj})$ behaviour is governed
by the end point behaviour of the distribution amplitude.
For a physical distribution amplitude that vanishes at the end points, the
obtained $(1-x_{Bj})$ power behaviour agrees with the counting rule result.
In the other case one finds a disagreement.

Moreover, to obtain the evolution in $Q$ we have included the renormalized
LC operator in the computation of the hard amplitude. Since the
initial condition of the renormalization group equation can be computed
perturbatively (this is not possible for the initial condition of the GLAP
equation) we have shown that the reference momentum square $Q_0^2$ is given
by the $x_{Bj}$ dependent factorization scale, i.e., by $m^2/(1-x_{Bj})$.

Note that the use of this treatment for QCD is straight forward.
It will be interesting to investigate the question of the $(1-x_{Bj})$
power behaviour on the dependence of the distribution amplitude and the
effects of higher twist terms.
It is to be expected that the founded value $Q_0^2 \sim m^2/(1-x_{Bj})$ will be
the same in QCD, so that the power behaviour of the valence quark distribution
functions will be changed [see Eq.\ (\ref{beh.ofDA})].

\section{Acknowledgments}

It is a pleasure for me to thank S. Brodsky for his suggestion to this
work and for stimulating discussions.
I am also indebted to B. Geyer and D. Robaschik for helpful discussions.

\end{document}